# Methanation of carbon dioxide over ceria-praseodymia promoted Ni-alumina catalysts. Influence of metal loading, promoter composition and alumina modifier

*Ali Lechkar[a], Adrián Barroso Bogeat[b,c,*], Ginesa Blanco[b,c], José María Pintado[b,c], Mohamed Soussi el Begrani[a]*

[a]*Département de Chimie, Equipe de Catalyse Environnementale, Faculté des Sciences, Université Abdelmalek Essaadi B.P. 2121 M'Hannech II, 93002-Tétouan, Morocco*

[b]*Departamento de Ciencia de los Materiales e Ingeniería Metalúrgica y Química Inorgánica, Facultad de Ciencias, Universidad de Cádiz, Campus Río San Pedro s/n, 11510 Puerto Real (Cádiz), Spain*

[c]*Instituto Universitario de Investigación en Microscopía Electrónica y Materiales (IMEYMAT), Facultad de Ciencias, Universidad de Cádiz, Campus Río San Pedro s/n, 11510 Puerto Real (Cádiz) Spain*

ABSTRACT

Two series of ceria-praseodymia promoted Ni-alumina catalysts were prepared from two different commercial modified alumina supports (3.5 wt.% $SiO_2$-$Al_2O_3$ and 4.0 wt.% $La_2O_3$-$Al_2O_3$) by the incipient wetness impregnation method in two successive steps. The resulting materials were characterized in terms of their physico-chemical properties by means of $N_2$ physical adsorption at -196 ºC, powder X-ray diffraction (XRD) and temperature programmed reduction with $H_2$ ($H_2$-TPR). Furthermore, the as-prepared catalysts were tested for the $CO_2$ methanation reaction in a fixed-bed reactor at atmospheric pressure, gas hourly space velocity (GHSV) of 72,000 $cm^3 \cdot (h \cdot g_{cat})^{-1}$ and $CO_2$/$H_2$ molar ratio of 1/4 over the temperature range from 25 up to 850 ºC. The influence of the nominal Ni loading (3, 5 and 10 wt.%), molar composition of the Ce/Pr mixed oxide promoter (80/20 and 60/40), and alumina modifier (silica and lanthana) on the catalytic performance was carefully analyzed. Among these three composition parameters, the alumina dopant and especially the Ni content appear to have by far a much more pronounced effect on both the $CO_2$ conversion and $CH_4$ selectivity as compared to the Ce/Pr mixed oxide composition. Specifically, from the catalytic tests the sample containing a 10 wt.% Ni loading, a Ce/Pr mixed oxide promoter of 80/20 molar composition, and silica as modifier provides the highest catalytic activity in terms of $CO_2$ conversion and $CH_4$ selectivity. Such behaviour has been ascribed to a complex interplay between several factors, mainly the larger fraction of catalytically active β-type NiO species and the lesser concentration of strong basic sites on the catalyst surface, as well as the electron back donation effect from the surface Ni atoms to the adsorbed $CO_x$ species, which favours the C–O bond cleavage (i.e., the rate-determining step of the methanation reaction). These findings are expected to be very helpful in order to rationally design synthetic strategies that allow developing highly active and low cost Ni-based catalysts for the $CO_2$ methanation reaction.

---

[*] Corresponding author at: Departamento de Ciencia de los Materiales e Ingeniería Metalúrgica y Química Inorgánica, Facultad de Ciencias, Universidad de Cádiz, Campus Río San Pedro s/n, 11510 Puerto Real (Cádiz), Spain
*E-mail adress*: adrian.barroso@uca.es (Adrián Barroso-Bogeat)

**1. Introduction**

Carbon-rich fossil fuels (i.e., petroleum, coal and natural gas) are expected to remain as the dominant worldwide source of energy through at least the next two decades [1]. This situation, together with the low efficiency of the energetic processes, has led to a steady and significant raise of the atmospheric levels of $CO_2$ [2-5], which is considered by far the major human-related greenhouse gas. In this regard, it should be highlighted that recent studies have estimated an excess of around 3.9% of $CO_2$ with respect to the natural "carbon cycle" [6,7]. Such an increase is generally agreed to be the primary reason for the rapid global warming observed during the past decades, with its inherent climate change and subsequent environmental disasters including sea level rise, desertification and extinction of species, among others [7,8]. Therefore, the reduction of the global $CO_2$ emissions is still an environmental issue of major concern and a real challenge for the international scientific community. As a result, intensive efforts are being devoted to the development of new technologies which allow reducing the buildup of $CO_2$ in the atmosphere. In principle, three main solutions have been proposed: (i) to reduce the amount of $CO_2$ produced, which requires both an improvement of energetic efficiency and a change in the primary energy source from C-rich fossil fuels to hydrogen and renewable energies, (ii) to capture $CO_2$ at its source and store it in the geological subsurface so that it will no longer be able to contribute to the global warming, and (iii) to promote the conversion of $CO_2$ into a variety of useful chemicals or fuels of high added value, such as methane, methanol, ethanol or dimethylether, by using appropriate catalysts [2,7-20].

Concerning the last strategy, $CO_2$ methanation (i.e., the reaction between $CO_2$ and $H_2$ yielding $CH_4$) emerges as the most advantageous reaction, since it is notably faster when compared to other hydrogenation reactions leading to the formation of hydrocarbons or alcohols [21]. Although the methanation reaction was first reported by Sabatier and Senderens in the early 1900s [22], it has only been applied on an industrial scale since 1970s, chiefly in the purification of synthesis gas for ammonia production.

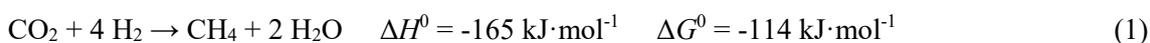

$CO_2 + 4\ H_2 \rightarrow CH_4 + 2\ H_2O \qquad \Delta H^0 = -165\ \text{kJ·mol}^{-1} \qquad \Delta G^0 = -114\ \text{kJ·mol}^{-1}$ (1)

As can be seen, the methanation of $CO_2$ is an exothermic and thermodynamically favourable reaction. Nevertheless, a complex process involving 8 electrons and affected by significant kinetic limitations is required in order to reduce the fully oxidized carbon atom to

$CH_4$. Moreover, the activation of the $CO_2$ molecule at low temperature is very difficult because of its high chemical inertness and thermodynamic stability. Therefore, the use of a suitable catalyst is imperative to achieve acceptable reaction rates and selectivities to $CH_4$, especially at low temperatures [3].

A number of supported and unsupported noble and transition metal catalysts (i.e., Ru, Rh, Pd, Pt, Ir, Os, Fe, Co, Cu, etc.) have been tested for $CO_2$ methanation [13,23-27]. Among them, those based on Ni supported on a variety of metal oxides, including $TiO_2$, $SiO_2$, $Al_2O_3$, MgO, $ZrO_2$, $Y_2O_3$, $CeO_2$, and so on, are by far the most extensively investigated under widely varying experimental conditions [13,19,28-31]. The best performances in terms of catalytic activity and selectivity to $CH_4$ have been reported for $Ni/Al_2O_3$ catalysts, together with high chemical and physical stability, mechanical resistance and relatively low price [5,29,31-35]. However, the main drawback associated with such catalysts is their rapid deactivation due to the sintering of Ni nanoparticles and severe carbon deposition when operating in high temperature conditions [5,36]. A plausible strategy to overcome this shortcoming consists of the design and development of novel promoted Ni-based catalysts, combining high catalytic activity, selectivity and resistance against carbon deposition, by incorporating rare earth oxides, such as lanthana, ceria, samaria and praseodymia [37,38]. Thus, $CeO_2$ has been frequently proposed as structural and electronic promoter for Ni catalysts because, as unique virtues, it is able to confer the following beneficial effects: (i) improves the thermal stability of alumina [38], (ii) increases the dispersion of Ni on the support [40,41], and (iii) modifies the properties of Ni by means of strong metal-support interactions [36,42].

In the present work, from two commercial modified aluminas, two series of novel ceria-praseodymia promoted Ni-alumina catalysts were prepared by a simple incipient wetness impregnation method in two successive steps (i.e., the first cycle for ceria-praseodymia mixed oxide and the second one for Ni). Resulting samples were characterized in terms of their physico-chemical properties by several techniques and tested as catalysts for the $CO_2$ methanation reaction at atmospheric pressure and temperatures up to 850 ºC in a fixed-bed reactor. The effects of the alumina support, the Ce/Pr molar ratio of the promoter and the Ni loading on the catalyst properties and catalytic performance were comprehensively examined and discussed. It is noteworthy that, to the best of our knowledge, this is the first time that the use of ceria-praseodymia mixed oxides as promoters in Ni/alumina catalysts for the $CO_2$ methanation is reported.

## 2. Experimental

*2.1. Materials and reagents*

Two commercial silica- and lanthana-modified alumina samples kindly supplied by Grace Davison, as received without any further treatment, were used as catalyst supports. Silica and lanthana contents were 3.5 wt.% and 4.0 wt.%, respectively. Metal nitrates, which are readily soluble in water, were employed as precursors. These were $Ce(NO_3)_3 \cdot 6H_2O$, $Pr(NO_3)_3 \cdot 6H_2O$ and $Ni(NO_3)_2 \cdot 6H_2O$, all of them being of analytical grade (i.e., > 99.9%) and directly used without additional purification. Both $Ce(NO_3)_3 \cdot 6H_2O$ and $Pr(NO_3)_3 \cdot 6H_2O$ were purchased from Rhodia (France), whereas $Ni(NO_3)_2 \cdot 6H_2O$ was procured from BDH Chemicals (United Kingdom).

*2.2. Preparation of the catalysts*

The preparation of the ceria-praseodymia promoted Ni-alumina catalysts was carried out by the method of incipient wetness impregnation [43-46] in two consecutive steps, as briefly described below. First, two Ce/Pr mixed oxides, with Ce/Pr molar ratios of 80/20 and 60/40, were deposited over both alumina supports by impregnation in one cycle with an aqueous solution containing a mixture of $Ce(NO_3)_3 \cdot 6H_2O$ and $Pr(NO_3)_3 \cdot 6H_2O$ in the appropriate molar ratio. The final mixed oxide loading was set to 25 wt.% with respect to alumina in all cases. After impregnation, the solids were oven-dried at 100 ºC overnight, ground in an agate mortar, sieved, and then calcined in a muffle furnace with a heating rate of 10 ºC·min$^{-1}$ up to 500 ºC and maintained at this temperature for 6 h. Such calcination temperature was selected in order to ensure the complete thermal decomposition of nitrates to metal oxides after impregnation and oven-drying, as previously demonstrated elsewhere [44]. The final composition of the resulting alumina-supported Ce/Pr mixed oxide systems was: 25% $Ce_{0.8}Pr_{0.2}O_{2-\delta}/Al_2O_3$-$SiO_2$, 25% $Ce_{0.6}Pr_{0.4}O_{2-\delta}/Al_2O_3$-$SiO_2$, 25% $Ce_{0.8}Pr_{0.2}O_{2-\delta}/Al_2O_3$-$La_2O_3$, and 25% $Ce_{0.6}Pr_{0.4}O_{2-\delta}/Al_2O_3$-$La_2O_3$. Herein, they will henceforth be referred to as CP(80/20)/Al-Si, CP(60/40)/Al-Si, CP(80/20)/Al-La, and CP(60/40)/Al-La, respectively. The final Ni catalysts were prepared by impregnation of the alumina-supported Ce/Pr mixed oxide samples with an aqueous solution containing the required $Ni(NO_3)_2 \cdot 6H_2O$ amount to achieve a nominal Ni loading of 3, 5 and 10 wt.% in the final products after one impregnation cycle. Subsequently, the impregnated samples in a series of successive steps were oven-dried at 100 ºC for 24 h, ground, sieved and finally calcined in a muffle furnace with a heating rate of 10 ºC·min$^{-1}$ up to 500 ºC and held at such temperature for 2 h.

As a whole, a broadly varied series of catalysts were prepared depending on the nature of the dopant introduced to stabilize the alumina support, the Ce/Pr molar ratio in the Ce/Pr mixed oxide promoter, and the nominal Ni loading. The codes assigned to the as-prepared catalysts are listed in Table 1.

**Table 1.** Preparation of the catalysts. Sample codes

| Support | Ce/Pr molar ratio | Nominal Ni loading / wt.% | Code |
|---|---|---|---|
| Al$_2$O$_3$-3.5 wt.% SiO$_2$ | 80/20 | 3 | Ni(3)/CP(80/20)/Al-Si |
| | | 5 | Ni(5)/CP(80/20)/Al-Si |
| | | 10 | Ni(10)/CP(80/20)/Al-Si |
| | 60/40 | 5 | Ni(5)/CP(60/40)/Al-Si |
| | | 10 | Ni(10)/CP(60/40)/Al-Si |
| Al$_2$O$_3$-4.0 wt.% La$_2$O$_3$ | 80/20 | 3 | Ni(3)/CP(80/20)/Al-La |
| | | 5 | Ni(5)/CP(80/20)/Al-La |
| | | 10 | Ni(10)/CP(80/20)/Al-La |
| | 60/40 | 5 | Ni(5)/CP(60/40)/Al-La |
| | | 10 | Ni(10)/CP(60/40)/Al-La |

*2.3. Characterization of the catalysts*

Textural characterization of the fresh catalysts was performed by physical adsorption of N$_2$ at -196 ºC using an automatic Autosorb iQ$_3$ equipment (Quantachrome) and working with relative pressures ($p/p^0$) in the range between 0.01 to 1.0. Before effecting the adsorption-desorption measurements, about 100 mg of sample was outgassed under vacuum at 200 ºC for 2 h in order to remove moisture and other gases from the laboratory atmosphere adsorbed on the sample surface. The measured N$_2$ adsorption-desorption isotherms provided valuable information regarding the pore size distribution in the micro- and mesoporosity regions, surface area and pore volume. The pore size distribution curves in the mesopore range were estimated by applying the Barrett-Joyner-Halenda (BJH) method [47] to the desorption branch of the isotherms. The apparent surface areas ($S_{BET}$) were calculated by means of the Brunauer, Emmett and Teller (BET) equation [48], which as a rule was applied in the $p/p^0$ range from 0.05 to 0.20. Furthermore, the total pore volumes ($V_p$) were derived from the volumes of N$_2$ adsorbed at $p/p^0$ = 0.99, expressed as liquid volumes.

Powder X-ray diffraction (XRD) patterns were collected for the alumina-supported Ce/Pr mixed oxide samples as well as for both the fresh and spent Ni catalysts at room temperature in a PANalytical X'Pert PRO diffractometer, operating with Ni-filtered Cu K$\alpha$ radiation ($\lambda$ = 1.5406 Å) at 40 kV and 40 mA. The specific acquisition conditions were: $2\theta$ range from 10º to 90º, step size 0.06º, and step counting time 120 s. Crystalline phases present in the samples were identified and indexed by comparing peak positions and intensities with JCPDS standard cards and with data previously reported in the literature.

The redox behaviour both of the alumina-supported Ce/Pr mixed oxide samples and the corresponding Ni catalysts was characterized by temperature-programmed reduction (TPR) followed by mass spectrometry (MS). TPR-MS experiments were conducted in a conventional experimental device coupled to a quadrupole mass spectrometer (Thermostar GSD301T1, Pfeiffer Vacuum). The amount of sample typically employed in each run was around 200 mg. Prior to starting the reduction experiments, all the samples were subjected to a standard cleaning pretreatment consisting of oxidation in a 60 cm$^3$·min$^{-1}$ STP flow of $O_2$(5%)/He at 500 ºC for 1 h, followed by slow cool down under the same oxidizing gas mixture to 150 ºC and then the flow was switched to pure He for further cooling down to room temperature. After this pretreatment, the TPR-MS analyses were performed in a flow of 60 cm$^3$·min$^{-1}$ STP of $H_2$(5%)/Ar from room temperature up to 950 ºC at a heating rate of 10 ºC·min$^{-1}$. The samples were held at this maximum temperature for 1 h. The main mass/charge (m/z) ratios recorded during the above experiments were 2 ($H_2^+$) to monitor the $H_2$ consumption and 18 ($H_2O^+$) for the concomitant formation of water.

*2.4. Catalytic activity tests*

$CO_2$ methanation experiments were carried out at atmospheric pressure in a U-shaped fixed-bed continuous flow quartz reactor loaded with a mixture of about 50 mg of catalyst and 100 mg of SiC as diluent, which was held by quartz wool. The incorporation of SiC was primarily aimed at avoiding the generation of hot spots along the catalyst bed during the catalytic tests. This reactor was introduced inside a vertical tubular electric furnace and the hot junction of a K-type thermocouple was placed close to the central part of the catalyst bed in order to accurately control the temperature in situ during the pretreatment and reaction. Prior to starting the catalytic essays, the catalysts were subjected to a reductive pretreatment in a 60 cm$^3$·min$^{-1}$ STP flow of $H_2$(5%)/Ar at 700 ºC for 30 min with a heating ramp of 10 ºC·min$^{-1}$, followed by purging with a 60 cm$^3$·min$^{-1}$ STP flow of pure Ar at the same temperature for 10 min to remove the chemisorbed $H_2$. Such pretreatment temperature was selected in order to ensure the complete reduction of the supported $Ni^{2+}$ species to metallic Ni while avoiding the formation of undesirable lanthanide aluminate phases with perovskite-type structure, as will be comprehensively discussed in a later section. Afterwards, the system was allowed to cool down to room temperature under the same Ar flow and then a gaseous mixture of $CO_2$(5%)/He and $H_2$(5%)/Ar with a molar ratio of 1/4 was fed to the reactor at a total flow rate of 60 cm$^3$·min$^{-1}$ STP. The methanation reaction was performed from room temperature up to 850 ºC with a heating rate of 10 ºC·min$^{-1}$ and a gas hourly space velocity (GHSV) of 72,000 cm$^3$·(h·g$_{cat}$)$^{-1}$. Such a maximum temperature was maintained for 30 min before cooling down to room temperature under the same reactant mixture flow. Effluent stream from the reactor was analyzed and the concentration of the gases of interest was quantified on line by a quadrupole mass spectrometer (Prisma QME-200-D, Pfeiffer Vacuum). The m/z ratios corresponding to the mass fragments of $H_2$, He, $CH_4$, $H_2O$, CO, and $CO_2$ were registered. $CO_2$ conversion

(abbreviated as $X_{CO2}$) and CH$_4$ selectivity and yield (denoted as $S_{CH4}$ and $Y_{CH4}$, respectively) were calculated by applying the following expressions:

$$X_{CO_2}(\%) = \left(\frac{Q_{CO_{2,in}} - Q_{CO_{2,out}}}{Q_{CO_{2,in}}}\right) \cdot 100 \tag{2}$$

$$S_{CH_4}(\%) = \left(\frac{Q_{CH_{4,out}}}{Q_{CO_{2,in}} - Q_{CO_{2,out}}}\right) \cdot 100 \tag{3}$$

$$Y_{CH_4}(\%) = \left(\frac{Q_{CH_{4,out}}}{Q_{CO_{2,in}}}\right) \cdot 100 \tag{4}$$

where $Q_{CO2,in}$ represents the molar flow rate (mol·min$^{-1}$) of CO$_2$ in the inlet stream, whereas $Q_{CO2,out}$ and $Q_{CH4,out}$ stand for the molar flow rates of CO$_2$ and CH$_4$ in the outlet stream, respectively.

## 3. Results and discussion

### 3.1. Characterization of the fresh catalysts

The N$_2$ adsorption-desorption isotherms measured for the alumina-supported Ce/Pr mixed oxide samples are depicted together in Fig. 1(a) for comparison purposes. Such isotherms by their shape belong to type IV of the IUPAC classification system [49], so that these materials can be considered as essentially mesoporous solids. In this regard, it should be noted that the small adsorption of N$_2$ at very low $p/p^0$ values (i.e., $p/p^0$ < 0.1) confirms the absence of any significant microporosity in the materials. Moreover, the steady but non-regular increase of the N$_2$ adsorption with the gradual rise of $p/p^0$ in the range from 0.1-0.2 to 0.8-0.9 as well as the much more pronounced increase at higher $p/p^0$ values close to 1 indicate that these samples posses a relatively broad pore size distribution in the region of mesoporosity. Furthermore, all isotherms exhibit well defined hysteresis loops in the desorption branches, associated with capillary condensation phenomena in mesopores wider than ~ 4 nm [49,50]. According to the recently updated IUPAC classification, the hysteresis loops closely resemble type H3, which is characteristic of non-rigid aggregates of plate-like particles giving rise to slit-shaped pores [49]. For the sake of brevity, the N$_2$ isotherms measured for the Ni catalysts have been omitted in the above figure since they are very similarly shaped to those registered for the alumina-supported Ce/Pr mixed oxide systems; the only noticeable change affected the amount of adsorbed N$_2$ as a result of the different surface area of the samples. Therefore, it follows that the preparation process of the Ni catalysts by impregnation and subsequent calcination did not bring about relevant modifications in the pore size distribution of the alumina-supported Ce/Pr oxides. This conclusion was corroborated from the pore size distributions estimated both for the supports and the Ni catalysts by applying the BJH method to the desorption branch of the corresponding isotherms. Such curves for selected samples are shown in Fig. 1(b). The only maximum centred at around 10 nm pore diameter in the plots denotes that the porosity distribution was monomodal for all the samples, irrespective of the commercial modified-alumina employed in their preparation. However, the distribution was slightly wider for the series of samples prepared from Si-doped alumina as compared to those obtained from La-doped alumina. Also notice that the peak is somewhat less intense and narrower for the Ni catalysts than for their corresponding supports, the effect being noticeably stronger with increasing Ni loading in the catalyst. This behaviour has been attributed to a certain blockage of the porosity of the support caused by the deposition of Ni particles, as previously reported for other lanthanide-promoted Ni-alumina catalysts [38].

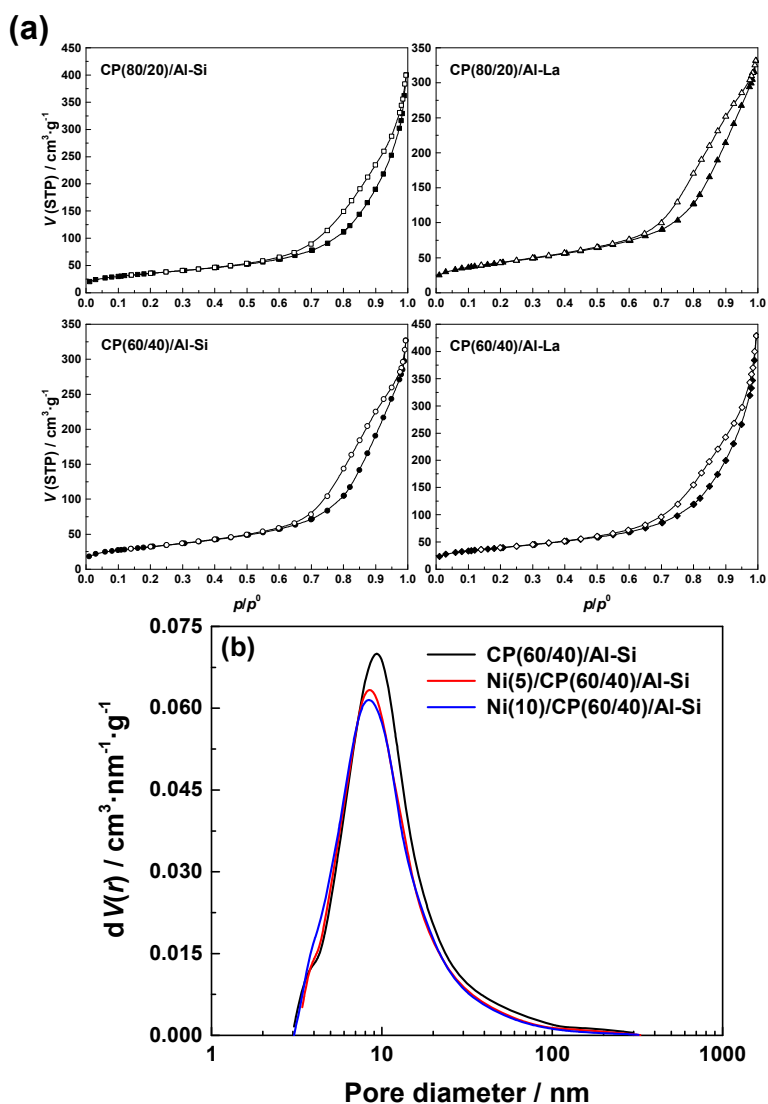

**Fig. 1.** Nitrogen adsorption/desorption isotherms at -196 ºC of alumina-supported Ce/Pr mixed oxides (a) and pore size distributions of selected Ni catalysts (b).

The textural data obtained for the alumina-supported Ce/Pr mixed oxide systems and the respective Ni catalysts are gathered in Table 2. As expected, for most of the prepared catalyst samples the incorporation of Ni led to a slight decrease both in the specific surface area and pore volume as compared to their respective supports, the reduction being obviously greater for those catalysts containing a higher Ni amount. In this connection, the specific surface area varied between ca. 132 and 153 $m^2 \cdot g^{-1}$ for the Si-doped catalysts, and in the narrower range from ca. 131 to 142 $m^2 \cdot g^{-1}$ for their La-modified counterparts.

**Table 2.** Textural data for the prepared alumina-supported Ce/Pr mixed oxides and Ni catalysts.

| Sample | $S_{BET}$ / $m^2 \cdot g^{-1}$ | $V_p$ / $cm^3 \cdot g^{-1}$ |
|---|---|---|
| CP(80/20)/Al-Si | 128.6 | 0.62 |
| Ni(3)/CP(80/20)/Al-Si | 153.2 | 0.59 |
| Ni(5)/CP(80/20)/Al-Si | 142.9 | 0.58 |
| Ni(10)/CP(80/20)/Al-Si | 136.4 | 0.57 |
| CP(60/40)/Al-Si | 141.9 | 0.66 |
| Ni(5)/CP(60/40)/Al-Si | 131.9 | 0.54 |
| Ni(10)/CP(60/40)/Al-Si | 136.1 | 0.51 |
| CP(80/20)/Al-La | 154.5 | 0.51 |
| Ni(3)/CP(80/20)/Al-La | 142.0 | 0.49 |
| Ni(5)/CP(80/20)/Al-La | 131.1 | 0.45 |
| Ni(10)/CP(80/20)/Al-La | 140.3 | 0.45 |
| CP(60/40)/Al-La | 115.7 | 0.51 |
| Ni(5)/CP(60/40)/Al-La | 141.3 | 0.50 |
| Ni(10)/CP(60/40)/Al-La | 131.9 | 0.45 |

The structural characterization of both series of alumina-supported Ce/Pr mixed oxide systems and Ni catalysts was accomplished by powder XRD. The recorded diffractograms are displayed in Fig. 2. As can be seen from this figure, they are all very similarly shaped regardless of the alumina dopant, being by far dominated by the characteristic reflections attributable to the cubic fluorite-type structure (space group $Fm$-$3m$) typical of ceria and ceria-praseodymia mixed oxides. The assignment of these peaks to a single or several $Ce_xPr_{1-x}O_{2-\delta}$ crystalline phases is an extremely complex task in view of the well known tendency of Pr to form a variety of stoichiometric and non-stoichiometric sub-oxides represented by the general formula $PrO_x$ with $x \leq 2$, whose diffraction patterns only differ slightly from that for pure ceria [51]. In this regard, the Ce/Pr mixed oxide peaks are asymmetric, with shoulders appearing at the left side of the main peak, thus suggesting the presence in the samples of at least two segregated fluorite-type phases with different cerium contents. One of them can be assigned to a cerium-rich mixed oxide giving rise to the main reflections observed in the diagrams at $2\theta$ values close to those of pure ceria, whereas the other one can be identified as a praseodymium-rich phase (or even pure praseodymia) related to the aforesaid shoulders. The shift of the peaks of this praseodymium-rich mixed oxide toward somewhat lower $2\theta$ values with regard to the position of pure ceria peaks is ascribable to a significant enlargement of the lattice parameter of the latter oxide due to the incorporation of some $Pr^{3+}$ cations, which are notably bigger than $Ce^{4+}$ cations (i.e., cationic radii were estimated to be 0.113 nm for $Pr^{3+}$ and 0.097 nm for $Ce^{4+}$ [43,52]). These results seem to be well in agreement with those previously found for identical alumina-supported Ce/Pr mixed oxides by means of both Rietveld analysis and advanced electron microscopy techniques [43,45]. These works also revealed that the spatial distribution of the praseodymium-rich phase was markedly different depending on the alumina modifier: as nanosized particles in the samples prepared from the Si-doped alumina and as a highly dispersed phase in the La-modified alumina samples. Furthermore, it should be noted that the diffraction peaks asymmetry and thereby the segregation of cerium-rich and poor crystalline phases is more evident for the supported mixed oxides with a Ce/Pr molar ratio of 80/20. In addition to the well defined Ce/Pr mixed oxide diffraction peaks, other much less intense reflections associated with γ-alumina polymorph can also be distinguished in the XRD diagrams.

After the successive Ni impregnation and calcination steps, a few broad and very weak reflections ascribable to nickel oxide (NiO) with face-centred cubic structure (space group *Fm-3m*) at 37.2º, 43.2º, 62.8º and 79.4º as well as to nickel aluminate ($NiAl_2O_4$) with spinel-type structure (space group *Fd-3m*) at 37.0º, 44.0º and 65.5º can be observed in the XRD patterns of the resulting catalyst samples, thus suggesting that these Ni-containing phases as a rule are highly dispersed on the surface of the supports and exhibit very small crystallite sizes. As expected, such diffraction peaks become gradually more intense and sharper with increasing Ni loading in the catalysts from 3 to 10 wt.%, which is unambiguously connected with a growth of crystallite size for both Ni species as previously suggested from textural characterization data. The formation of the aluminate phase clearly indicates that the $Ni^{2+}$ species in the impregnation aqueous solution not only interacted intimately with the supported Ce/Pr mixed oxides but also with the modified-alumina substrates during the preparation process of the catalysts.

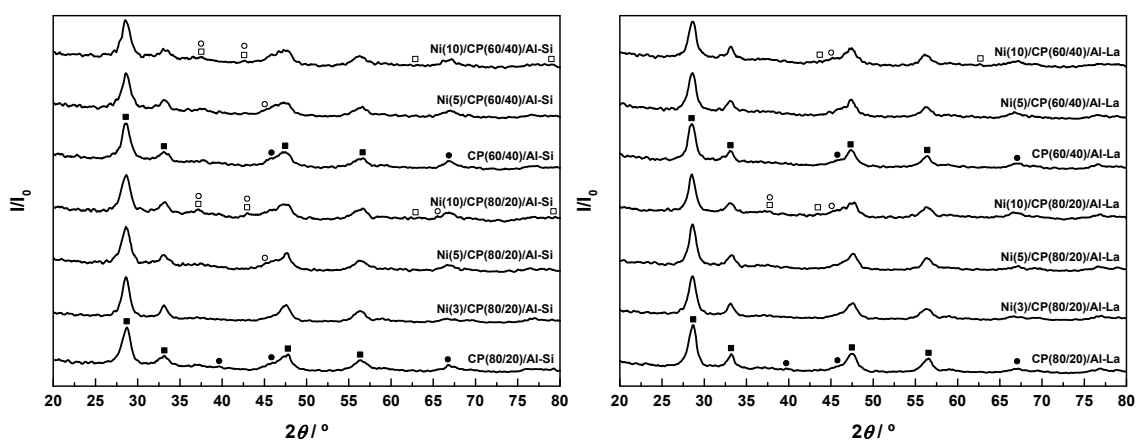

**Fig. 2.** Powder XRD patterns of alumina-supported Ce/Pr mixed oxides and fresh Ni catalysts. Caption: ■ $Ce_xPr_{1-x}O_{2-\delta}$ fluorite structure, ● γ-$Al_2O_3$, □ NiO, and ○ $NiAl_2O_4$.

The reduction behaviour of both the alumina-supported Ce/Pr mixed oxide systems and the corresponding Ni catalysts was studied by means of the TPR-MS technique in $H_2(5\%)$/Ar. For the sake of brevity, only the curves registered for the samples containing a 10 wt.% nominal Ni loading after the standard cleaning pretreatment in $O_2(5\%)$/He at 500 ºC for 1 h are plotted in Fig. 3, as they are considered to be representative of the overall redox behaviour of the prepared catalysts. As far as the diagrams for the alumina-supported Ce/Pr mixed oxides are concerned, readers are referred to our previous works [46,53]. In brief, both CP(80/20)/Al-Si and CP(80/20)/Al-La samples exhibit a quite similar behaviour at temperatures ranging from 300 to 800 ºC. For the former, the onset of water evolution occurs at 350 ºC with two maxima centred at around 400 and 580 ºC, while for the latter the reduction starts at a somewhat higher temperature, the first peak being located at 450 ºC and the second one also at 580 ºC. By comparing these TPR profiles with that obtained for a bulk mixed oxide prepared with identical Ce/Pr molar ratio [53,54], the first maximum could be assigned to the reduction of most $Pr^{4+}$, whereas the second peak could be mainly attributed to the reduction of $Ce^{4+}$ [55,56]. In addition to the reduction of the supported Ce/Pr mixed oxides, the evolved water coming from the dehydroxylation of the alumina supports should also be borne in mind. Indeed, the release of a small amount of water between 660 and 800 ºC has been evidenced during the TPR runs carried out for both alumina substrates under the same conditions as for their respective supported Ce/Pr mixed oxides [53], likely as a result of the removal of residual hydroxyl groups remaining in the samples after the cleaning pretreatment. Finally, other prominent features of the curves are the sharp and intense peaks centred at about 940 ºC, which have been connected with the bulk reduction of the supported mixed oxides and the subsequent incorporation of the $Ce^{3+}$ and/or $Pr^{3+}$ cations into the alumina matrix to yield a lanthanide aluminate phase ($LnAlO_3$, Ln standing for $Ce^{3+}$ and/or $Pr^{3+}$) with perovskite-type structure (space group *Pm*-3*m*) [45,46,53,57-59]. In view of the relative intensity of the aforesaid peaks, the extent of the aluminate formation seems to be much greater for the La-modified samples as compared to those containing Si, which may act as a barrier against diffusion of the lanthanide cations into the alumina lattice [45]. Regarding the Ni catalysts, a number of processes involving the reduction of both the supported Ni phases and Ce/Pr mixed oxides (observed as $H_2$ consumption accompanied by the concomitant water evolution), as well as the alumina dehydroxylation (detected only as water release), occur along the entire temperature range, thus leading to the complex TPR profiles shown in Fig. 3. Such complexity is exacerbated by the fact that several reducible NiO species giving rise to reduction peaks at characteristic temperatures are likely to be present in the fresh catalysts depending on the nature of their interaction with the oxide support [60,61]. Furthermore, it is also worth highlighting that the aforesaid NiO species may be supported both on the Ce/Pr mixed oxides and the modified aluminas, thus making the interpretation of the TPR curves an even more difficult task. According to the most common interpretation found in the literature, three different reducible NiO species have been identified in supported Ni catalysts: α, β and γ [62-64]. The reduction peak associated with α-type NiO appears between 460 and 500 ºC and has been usually assigned to surface amorphous species or to very fine reducible species, which weakly interacted with the support. Thus, the greatly sloped shoulder at around 470 ºC displayed by the TPR profile for the Ni(10)/CP(80/20)/Al-La catalyst may be indicative of the presence in the sample of this kind of species. In this connection, it should be pointed out that the reduction of α-NiO has been suggested to promote the formation of large size Ni particles [65], which would account for the noticeable increase in the average crystallite size (as estimated from the respective XRD patterns by the Scherrer equation) of metallic Ni with respect to that of NiO for all the prepared catalysts after their reduction at 700 ºC. β-type NiO brings about a prominent and relatively broad reduction band at mid temperatures with a maximum typically centred in the range from 600 to 700 ºC, being related to NiO species in a stronger interaction degree with the support than α-type NiO ones [65,66]. The clearly visible maxima at around 620 and 650 ºC in the TPR traces for the Ni(10)/CP(80/20)/Al-La and Ni(10)/CP(80/20)/Al-Si catalysts, respectively, seem to indicate that these β-type NiO species are the predominant among those containing Ni and present in the samples. Finally, the shoulders registered in the vicinity of 800 ºC may be ascribed to the reduction of $NiAl_2O_4$ (i.e., the so-called γ-type NiO) [62,65,67], whose existence in the calcined catalysts in the form of well dispersed and very small crystallites had been previously inferred from their XRD patterns.

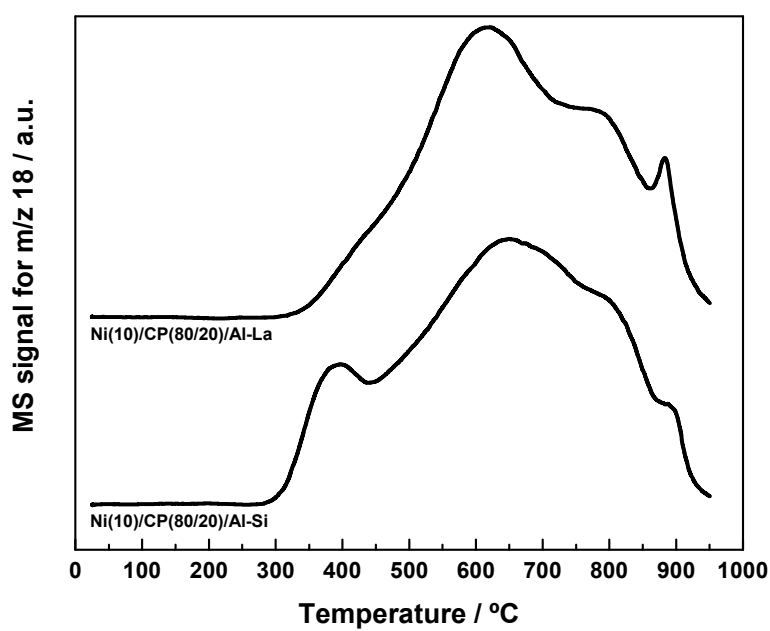

**Fig. 3.** TPR-MS profiles of representative fresh Ni catalysts calcined at 500 ºC.

Metallic Ni is generally agreed to be the catalytic active phase for $CO_2$ methanation [31,68,69], so it becomes evident that the catalysts prepared in the present work must be subjected to a reduction pretreatment prior to performing the catalytic tests [70] in order to transform the supported NiO and NiAl$_2$O$_4$ species into metallic Ni. Nevertheless, it should be taken into account that the alumina-supported Ce/Pr mixed oxides undergo significant structural changes during the reduction treatments, which ultimately give rise to the formation of a lanthanide aluminate phase (LnAlO$_3$, Ln standing for La$^{3+}$, Ce$^{3+}$ and/or Pr$^{3+}$) with perovskite-type structure (space group *Pm*-3*m*). Such phase is well known to be very stable, thus resulting in a dramatic redox deactivation of the supported mixed oxide due to the blockage of the reducible Ce and Pr cations in the +3 oxidation state [45,46]. According to our previous works [45,46], the temperature at which the formation of the undesirable aluminate phase starts during the reduction treatment strongly depends on the nature of the alumina modifier, being of around 800 ºC for the La-doped samples and 900 ºC for the Si-modified ones. In addition to this, high reduction temperatures usually cause sintering of the Ni (nano)particles supported on the catalyst surface, leading to a marked loss of active surface area and thereby to a reduction of the activity and even to the total deactivation of the catalyst [70]. Therefore, the temperature of the reduction pretreatment should be high enough to ensure the complete reduction of the different Ni$^{2+}$ species present in the catalyst samples to metallic Ni, but also low enough to avoid the formation of the perovskite-like phase with its inherent loss of the redox properties of the Ce/Pr mixed oxides, as well as the sintering of the active Ni (nano)particles. Based on the TPR profiles shown in Fig. 3, a temperature of 700 ºC has been selected to carry out the reduction pretreatment of the fresh catalysts. In order to check both the absence of the lanthanide aluminate and the presence of metallic Ni in the resulting prereduced samples, their XRD patterns were acquired (not displayed here for the sake of brevity). From them, the reduction of Ni$^{2+}$ species is confirmed by the appearance of reflection peaks ascribable to metallic Ni with face-centred cubic structure (space group *Fm*-3*m*), while no evidence of the formation of the perovskite phase is found at the detection level of the XRD technique. Finally, it should be highlighted that both the catalytic activity and selectivity of Ni-based catalysts for CO$_2$ methanation are largely influenced by the reduction pretreatment [70]. In this regard, it has been reported that for Ni-alumina catalysts low reduction temperatures as a rule favour the production of higher hydrocarbons, whereas higher temperatures are associated with a greater activity and selectivity towards CH$_4$ [71].

*3.2. Catalytic activity*

Fig. 4 depicts the results of the catalytic activity tests for the as-prepared Ni catalysts in the reaction of CO$_2$ methanation over the temperature range from 200 up to 600 ºC (i.e., data registered in the temperature ranges from 25 to 200 ºC and above 600 ºC have been omitted in the plots due to both the CH$_4$ yield and selectivity are almost negligible). From these figures, it becomes apparent that all the catalysts exhibit a similar behaviour in terms of CH$_4$ yield, CO$_2$ conversion and $CH_4$ selectivity, being completely inactive at temperatures below 200 ºC. As far as $CH_4$ yield is concerned (see Fig. 4(a)), the traces closely resemble a Gaussian-type curve (i.e., volcano-shaped trend), with maxima as a rule centred at 350 ºC for Si-doped catalysts and at somewhat higher temperatures around 400 ºC for their La-modified counterparts. Such a profile consisting of an increase of $CH_4$ yield with temperature until a maximum is reached, followed by a drop at greater temperatures, is a direct consequence both of the thermodynamics and kinetics of the process. It is well known that $CO_2$ methanation is a strongly exothermic reaction, which renders it unfavourable under high temperature conditions leading to very low $CH_4$ yield values. Furthermore, the reverse water gas shift reaction (rWGS, eq. (5)), which is an endothermic process, starts to dominate over the methanation reaction in this temperature range, as clearly seen from the CO yield plots gathered in Figs. 4(b) and (c).

$CO_2 + H_2 \leftrightarrow CO + H_2O \quad \Delta H^0 = 41 \text{ kJ·mol}^{-1}$ 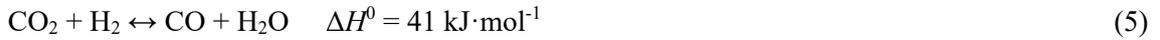 (5)

This predominance of the rWGS reaction results in the preferential formation of CO instead of $CH_4$ and further reduces the $CH_4$ yield [20,69,72,73]. On the other hand, the $CO_2$ methanation is a very complex process involving an 8-electron reduction and thereby affected by severe kinetic limitations, so that in addition to the presence of the catalyst a certain temperature is required to achieve significant amounts of $CH_4$, giving rise to the ascending branch of the curves [38].

Regarding the $CO_2$ conversion, the plots in Figs. 4(d) and (e) show a steady but non-regular increase over the whole investigated temperature range, with two sections of markedly different slope being clearly observed. The first portion extends from 200 ºC up to a temperature close to that at which the maximum of the $CH_4$ yield curve is reached (i.e., around 350 ºC for Si-modified catalysts and 400 ºC for La-doped ones) and the $CO_2$ conversion is almost exclusively attributable to the $CO_2$ methanation, the contribution of rWGS reaction at such temperatures being very low due to its endothermic character, as can be deduced from Figs. 4(b) and (c)). Specifically, the CO yield as a rule is estimated to be below 10% at 350 ºC for all the tested samples and slightly higher for La-modified catalysts as compared to those containing Si. Moreover, at this temperature the catalytic performance is observed to increase in the following order: Ni(3)/CP(80/20)/Al-La < Ni(5)/CP(60/40)/Al-La < Ni(3)/CP(80/20)/Al-Si ≈ Ni(5)/CP(80/20)/Al-La < Ni(5)/CP(60/40)/Al-Si < Ni(10)/CP(80/20)/Al-La ≈ Ni(10)/CP(60/40)/Al-La < Ni(5)/CP(80/20)/Al-Si < Ni(10)/CP(60/40)/Al-Si < Ni(10)/CP(80/20)/Al-Si. By contrast, the second section comprises temperatures above 350-400 ºC, where the conversion of $CO_2$ into CO through the rWGS reaction becomes dominant while the production of $CH_4$ starts to decline. In fact, from Figs. 4(b) and (c) it is evident that beyond 550 ºC the $CO_2$ conversion may be entirely ascribed to the aforementioned reaction.

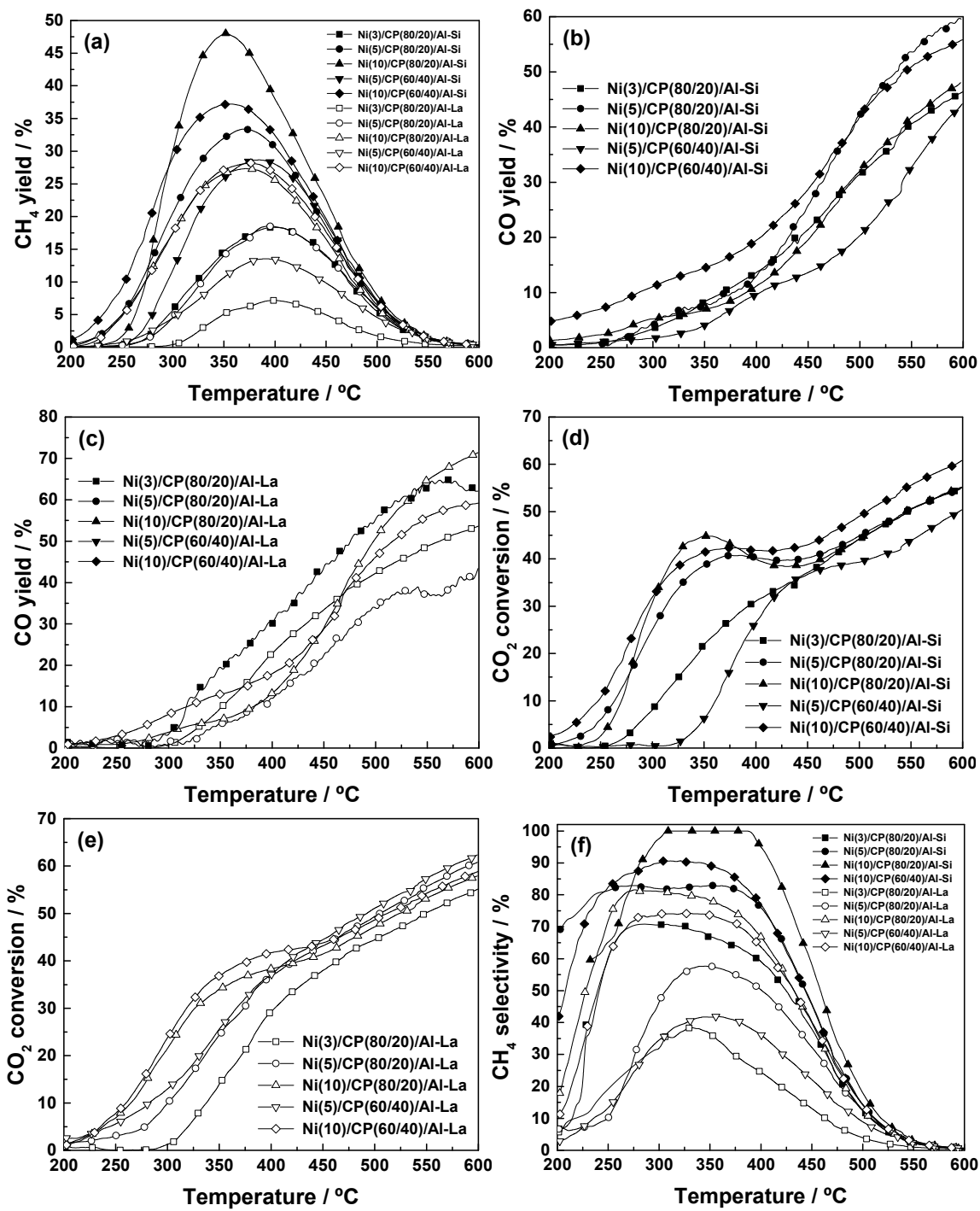

**Fig. 4.** Catalytic performance of the fresh Ni catalysts in the $CO_2$ methanation at atmospheric pressure, GHSV of 72,000 $cm^3 \cdot (h \cdot g_{cat})^{-1}$ and $CO_2/H_2$ molar ratio of 1/4 over the temperature range from 200 to 600 ºC. (a) $CH_4$ yield, (b) CO yield for Si-doped catalysts, (c) CO yield for La-doped catalysts, (d) $CO_2$ conversion for Si-doped catalysts, (e) $CO_2$ conversion for La-doped catalysts, and (f) $CH_4$ selectivity.

Finally, the $CH_4$ selectivity curves displayed in Fig. 4(f) are similarly shaped to those obtained for $CH_4$ yield. Nevertheless, the maxima of the former plots are shifted at lower temperatures, the extent of this change for each catalyst depending on the temperature at which the contribution of rWGS reaction begins to be significant (cf. Figs. 4(b) and (c)). In any case, the highest selectivity to $CH_4$ is attained at temperatures ranging from around 300 to 400 ºC for all the tested catalysts. Then, the $CH_4$ selectivity starts to decline sharply with the reaction temperature increase, being almost negligible at 600 ºC and above because of the predominance of the rWGS reaction.

An in-depth analysis of the catalytic activity data allows studying the influence of the Ni loading, the Ce/Pr molar ratio of the mixed oxide promoter and the alumina modifier on the catalytic performance of the prepared catalysts in the $CO_2$ methanation reaction. As stated earlier, the major contribution to $CO_2$ conversion over the temperature range from 200 to 450 ºC comes from the reaction of $CO_2$ methanation, so that the effect of the aforesaid three factors on the catalytic performance will be mainly evaluated hereafter in terms of $CO_2$ conversion, thereby obviating the contribution of the rWGS reaction.

*3.2.1. Effect of Ni loading*

From Figs. 4(d) and (e) it follows that over the aforesaid temperature range the $CO_2$ conversion markedly increases with Ni loading in the catalyst from 3 to 10 wt.%, irrespective of the molar composition of the Ce/Pr mixed oxide promoter and the nature of the alumina dopant. Since the pioneering work conducted by Kester et al. [74], a similar effect of metal content on the catalytic performance of Ni catalysts supported on a variety of materials, including zeolites [75], γ-alumina [66,76-78], mesoporous silica nanoparticles [79], titania [80], and so on, in the $CO_2$ methanation reaction has been extensively reported in the literature. There is general agreement that such behaviour can be largely attributed to an increase in the number of active sites in the surface of the catalysts with increasing Ni loading. Furthermore, the reducibility of the supported Ni-containing species has also been found to improve with the increment of metal content for Ni-alumina catalysts [74,81]. In this connection, based on TPR analyses at least two different reaction sites for $CO_2$ methanation were initially identified in these catalysts, (i) Ni crystallites coming from the reduction of NiO and (ii) Ni atoms surrounded by oxygen atoms from the alumina lattice, the latter being much lesser reactive than the former. According to more recent studies [81], the more reactive sites may be correlated with the more reducible α- and β-type NiO species, while the less reducible γ-NiO species are likely associated with the less reactive sites. Both sites are present in catalysts with Ni loadings ranging from 1.8 to 15 wt.% as a consequence of the different interactions between the $Ni^{2+}$ species and the alumina support during the preparation and the reduction pretreatment [74]; however, the fraction of the more reducible and thereby reactive species increases as the Ni loading on the catalysts does. Conversely, the less reducible and reactive γ-type NiO species become dominant with decreasing Ni content in the catalysts [81]. Finally, several authors have recently pointed out that β-type NiO species play a pivotal role as the main active sites for methanation reactions [81,82] because they can be easily reduced to relatively small-sized and highly active metallic Ni particles after reduction pretreatment [84]. Since the fraction of such species increases with the supported Ni content [65,81,83,84], it is expected that those catalysts containing a 10 wt.% metal loading exhibit the highest catalytic conversion.

In addition to the conversion, the selectivity toward $CH_4$ is also strongly influenced by the Ni loading. Thus, from Fig. 4(f) it becomes clear that between 200 and 450 ºC the $CH_4$ selectivity follows a similar trend to that noted above for the $CH_4$ yield and $CO_2$ conversion, i.e., the selectivity significantly increases with metal content. This behaviour may be explained by looking at the proposed mechanism for the $CO_2$ methanation over catalysts based on supported metals. There is some agreement that this reaction proceeds through the initial conversion of $CO_2$ into CO, the subsequent reaction following the same mechanism as CO methanation [85-88]. In the same way, this latter is also a two step reaction involving first the CO adsorption and dissociation at the surface of metal particles to yield adsorbed carbon and then its stepwise hydrogenation to $CH_4$. On the basis of both experimental data and theoretical calculations [88-91], the CO dissociation has been suggested as the rate-determining step of the process. In this connection, three different CO species adsorbed on Ni surface have been reported: (i) linear CO, (ii) bridged CO, and (iii) twin CO, whose activity towards the breaking of C–O bond varies according to the sequence: twin CO < linear CO < bridged CO, as inferred from the relative strength of the π back-bonding established between the antibonding π* orbitals of adsorbed CO molecules and the d orbitals of surface Ni atoms [92,93]. Consequently, it is expected that the more the bridged CO species formed on the catalyst surface, the greater the probability of the cleavage of C–O bond and thereby of the formation of surface carbon, which is subsequently hydrogenated to $CH_4$, and the lesser the probability of obtaining other hydrocarbons and oxygenates as by-products. Because the number of active bridged CO species is much greater for larger Ni particles [94], it is clear that the selectivity to $CH_4$ must be higher for those catalysts containing a 10 wt.% Ni loading.

*3.2.2. Effect of the composition of the Ce/Pr mixed oxide promoter*

Many efforts have been made to improve both the activity and stability of Ni-based catalysts for the $CO_2$ methanation reaction. As far as stability is concerned, one of the main challenges to be overcome is connected with the high tendency of these catalysts to suffer from coke deposition during the reaction, which leads to their rapid deactivation, especially when carbon deposits grow as filaments or nanotubes [95]. Therefore, the design and preparation of coke resistant Ni catalysts has become a research topic of major concern. The most widely employed strategy to achieve this goal consists of the incorporation of promoters in the catalyst formulations. Such promoters are able to enhance the coke resistance by increasing the adsorption of $CO_2$ and the rate of the surface reaction or by decreasing the rate and degree of $CH_4$ decomposition [95,96]. In fact, promoters with basic properties favour the reaction between $CO_2$ and carbon, thus improving the coke resistance of the catalysts. Similarly, promoters with redox properties have been extensively reported to increase the coke resistance. Finally, a strong metal-support interaction (SMSI) between the Ni and the promoter results in a similar effect. In view of these facts, a promoter based on Ce/Pr mixed oxides was selected in the present work because of these materials gather most of the abovementioned properties, thereby rendering coke resistance of the resulting catalysts easier.

From Fig. 4, it is clearly proven that the composition of the Ce/Pr mixed oxide promoter also influenced the catalytic activity for $CO_2$ methanation. In fact, for those catalysts containing the same nominal Ni loading and alumina modifier, the $CO_2$ conversion as well as the $CH_4$ yield and selectivity were found to improve with increasing Ce/Pr molar ratio from 60/40 to 80/20, this effect being more marked for higher metal contents. Notwithstanding, the influence was not so pronounced as compared to that observed for Ni loading. Such a very slight effect of the promoter composition on the catalytic performance of the prepared catalysts might be chiefly accounted for by the absence of remarkable differences in terms of electronic and chemical properties, especially of reducibility, between both supported lanthanide mixed oxides. This similarity had been previously pointed out by Borchert et al. [97] for nanostructured ceria-praseodymia samples with a 20 and 40 mol.% Pr content. By applying a number of characterization techniques, these authors concluded that the two samples were almost identical as far as their reduction behaviour and surface chemical composition (i.e., the atomic fractions of $Ce^{3+}$ and $Pr^{3+}$) are concerned. Furthermore, the minor differences in the catalytic activity could also be tentatively ascribed to the formation of stable carbonates on the surface of the mixed oxide promoters. As previously reported by Giménez-Mañogil et al. [98], both the amount and stability of carbonates dramatically increases with Pr content for $Ce_xPr_{1-x}O_{2-\delta}$ catalysts. Therefore, it would expected that after the reduction pretreatment the extent of the formation of stable carbonate-like species owing to the interaction of $CO_2$ molecules in the reactive feed stream with the catalyst samples at low reaction temperatures was higher for those containing the CP(60/40) promoter. Obviously, the $CO_2$ molecules forming these stable carbonates are much more difficult to hydrogenate to $CH_4$, thus inhibiting the methanation reaction and explaining the somewhat lower catalytic performance observed for the CP(60/40) catalysts.

On the other hand, it should be highlighted that the Ce/Pr mixed oxide promoters are likely to play a key role in the improvement of the catalytic activity as compared to conventional Ni-alumina catalysts. Thus, the surface $Ce^{3+}$ and $Pr^{3+}$ sites of the prereduced catalysts are able to promote the dissociation of C–O bond in the adsorbed CO species, which is the rate-limiting step of the methanation reaction, by enhancing the electron back donation process [92]. Such a beneficial effect can also be explained on the basis of the hard and soft acids and bases (HSAB) theory developed by Pearson [99]. According to it, $Ce^{3+}$ and $Pr^{3+}$ surface sites are classified as hard acids due to the essentially ionic character of their bonds [100], so they are expected to bind strongly to the oxygen atom of the adsorbed CO molecules, which is considered a strong base. As a consequence of these interactions, the C–O bond is notably weakened, thus making easier its cleavage and hence the formation of surface carbon and its ensuing hydrogenation to give $CH_4$. At this point, it is worth noting that the CO dissociation might be also favoured by the well-known oxygen storage capacity (OSC) of Ce/Pr mixed oxides [43-46,97], which allows them to effectively incorporate the oxygen atoms from the C–O bond breaking into their lattices due to the presence of a high number of oxygen vacancies induced by the insertion of Pr in the ceria crystalline structure.

*3.2.3. Effect of the alumina modifier*

Finally, the influence of the alumina modifier (i.e., silica and lanthana) on the catalytic performance of the prepared catalysts for the $CO_2$ methanation reaction was also analyzed in the present work. From Fig. 4, it is clearly noted that, for the same nominal Ni loading and composition of the Ce/Pr mixed oxide promoter, the catalyst samples prepared from Si-doped alumina show remarkably higher $CO_2$ conversion and $CH_4$ yield and selectivity when comparing to their counterparts supported on La-doped alumina. The origin of such an effect seems to be much clearer than that of the mixed oxide promoter composition and it has been mainly associated with two factors. The first one is the above-discussed formation of very stable lanthanide aluminate phases with perovskite-type structure during the pretreatment of the catalysts under reducing atmosphere at 700 ºC and the inherent redox deactivation of the Ce/Pr mixed oxides owing to the blockage of the $Ce^{3+}$ and $Pr^{3+}$ cations [45,46]. Despite the relatively low pretreatment temperature selected, it was impossible to entirely avoid the formation on the catalysts' surface of small amounts of the detrimental perovskite-like phase with very low crystallite size, which makes it undetectable for the powder XRD technique. In this regard, it is worth highlighting that the reduction pretreatment temperature of 700 ºC, even though markedly lower than those found from $H_2$-TPR analyses for the onset of the lanthanide aluminate formation for the La- and Si-modified aluminas (i.e., around 800 and 900 ºC, respectively), may be enough to promote this solid state reaction between the Ce/Pr mixed oxides and both alumina supports as a result of the prolonged soaking time of 1 h at such temperature. By taking into account the works previously carried out by our research group on these modified-alumina supported Ce/Pr mixed oxide systems [44-46], it may be concluded that the extent of the unwanted perovskite-like phase formation must be much greater for the La-doped catalyst samples as compared to the Si-modified ones. Therefore, the larger lanthanide aluminate content for the former catalysts leads to a higher redox deactivation of the Ce/Pr mixed oxide promoters, which redounds to their lower catalytic activities for the $CO_2$ methanation reaction. An in-depth analysis of the origin of the dissimilar effects of the two alumina dopants on the redox deactivation of the supported Ce/Pr mixed oxides was accomplished by means of a variety of techniques, including electron microscopy, XRD, FT-IR spectroscopy and XPS, and reported elsewhere [44-46]. Briefly, the obtained results unequivocally suggest that the $SiO_2$ modifier provides an effective physical barrier against the incorporation of the $Ce^{3+}$ and $Pr^{3+}$ cations into the alumina matrix, thus preventing the formation of large amounts of the redox inactive perovskite phase. On the contrary, $La_2O_3$ doping greatly increases the affinity of the alumina substrate towards the supported lanthanide elements, which renders their integration much easier with the consequent dramatic loss of redox properties of the Ce/Pr mixed oxides.

The second factor contributing to the better catalytic performance observed for the silica-containing catalysts in the $CO_2$ methanation reaction is the relative basicity of the adsorption active sites on the surface of the modified-alumina supports. According to the literature [101], the presence of moderate and weak basic sites on the support enhances the catalytic activity, while those of strong basic character are not involved in the methanation reaction due to the formation of highly stable carbonate species. Since γ-alumina mostly exhibits strong basic sites, it is expected that the incorporation of $SiO_2$, a well-known acid oxide, as dopant notably reduces both the population and strength of surface basic sites, thereby improving the catalytic performance. Regarding the $La_2O_3$ modifier, the opposite applies because of the acknowledged basic character of this metal oxide, which confers it a high reactivity against $CO_2$ [102]. Additionally, it should be pointed out that not only the $CO_2$ conversion and $CH_4$ yield but also the $CH_4$ selectivity are notably affected by the basicity of the modified-alumina supports (cf. Fig. 4(f)). Indeed, it has been recently reported that the selectivity to $CH_4$ is appreciably improved by decreasing both the concentration and strength of surface basic sites on the support [103], an observation which would account for the greater $CH_4$ selectivity obtained with the Si-doped catalyst samples in comparison with those containing La.

4. Conclusions

From two commercial modified alumina supports (3.5 wt.% $SiO_2$-$Al_2O_3$ and 4.0 wt.% $La_2O_3$-$Al_2O_3$), the preparation of two series of ceria-praseodymia promoted Ni-alumina catalysts was accomplished by the incipient wetness impregnation method in two successive steps: (i) impregnation of the alumina support with the promoter precursor, and (ii) impregnation of the resulting alumina-supported ceria-praseodymia system with the Ni precursor. The as-prepared catalyst samples were first characterized in terms of their physico-chemical features by $N_2$ physical adsorption, powder XRD and $H_2$-TPR and then studied for the reaction of $CO_2$ methanation. Special emphasis is paid to the influence on the catalytic performance of the nominal Ni loading (3, 5 and 10 wt.%), molar composition of the Ce/Pr mixed oxide promoter (80/20 and 60/40), and alumina modifier ($SiO_2$ and $La_2O_3$). The obtained results allow drawing the following main conclusions:

1. Among the three investigated composition parameters, metal content shows by far the more pronounced influence, followed by the nature of the alumina dopant. In stark contrast, the effect of the Ce to Pr molar ratio seems to be almost negligible.
2. Both the amount and fraction of β-type NiO species, which are widely considered as the key active sites for methanation, markedly increase with Ni content up to 10 wt.%.
3. The ceria-praseodymia promoter favours the cleavage of the C–O bond, which is the rate-limiting step of the methanation reaction, by enhancing the electron back donation from the surface Ni atoms to the adsorbed $CO_x$ species.
4. The $SiO_2$ dopant significantly decreases both the extent of the formation of the redox inactive lanthanide aluminate with perovskite-type structure and the strength of the basic sites on the catalyst surface, whereas the opposite applies to the $La_2O_3$ modifier.
5. As a result of the complex balance between the three aforesaid factors, the catalyst formulation containing a 10 wt.% Ni loading, a Ce/Pr molar ratio of 80/20 and $SiO_2$ dopant has been evidenced as the best in terms not only of $CO_2$ conversion but also of $CH_4$ yield and selectivity at low temperatures.

**Acknowledgements**

Financial support from MINECO/FEDER (Project MAT2013-40823-R) and Junta de Andalucía (Groups FQM-110 and FQM-334) is gratefully acknowledged. Adrián Barroso-Bogeat thanks support from the "Juan de la Cierva-Formación" Fellowship Program of MINECO (FJCI-2015-25999).